\documentstyle{article}
\def\ie{{\rm i.e.\/}\ }

\def\etc{{\rm etc\/}\ }

\def\bar{\overline}
\def\ZZ{\mbox{\rm Z}\hskip-4pt \mbox{\rm Z}}

\def\CC{\mbox{\rm C}\hskip-6pt\mbox{l} \;}
\def\id{\mbox{\rm 1}\hskip-3pt \mbox{\rm l}}

\newcommand{\bg}{\begin{equation}}
\newcommand{\eg}{\end{equation}}
\def\su21{$SU(2\, \vert \, 1)$}

\font\ten=cmbx10 at 12pt

\begin{document}

\begin{titlepage}

\begin{center}

\renewcommand{\thefootnote}{\fnsymbol{footnote}}

{\ten Centre de Physique Th\'eorique - CNRS - Luminy, Case
907}
{\ten F-13288 Marseille Cedex 9 - France }

\vspace{1,5 cm}

{\Large Comments about Higgs fields, noncommutative geometry and the
standard model$^{\ast )\ast \ast )}$}

\bigskip

G. Cammarata \\[0,3cm] R. Coquereaux\\

\end{center}

\vspace{2cm}

\begin{abstract} We make a short review of the formalism that
describes Higgs and Yang Mills fields as two particular cases of an
appropriate generalization of the notion of connection. We also
comment about the several variants of this formalism, their interest,
the relations with noncommutative geometry, the existence (or lack of
existence) of phenomenological predictions, the relation with Lie
super-algebras etc.

\end{abstract}

\vspace{2 cm}

\noindent anonymous ftp or gopher: cpt.univ-mrs.fr
 \vspace{2 cm}


\vfill {\small  $^{\ast )}$ Work supported in part by a PROCOPE
project  between Mainz University and CPT Marseille-Luminy. }\par
{\small $^{\ast \ast)}$ Based in part on lectures given by R.C. at
the   VI Simposio Argentino de F\'isica Te\'orica de Particulas y
Campos,  Bariloche (Argentina, January 95) and at the Schladming
Winter School (Austria, March 95)}\\ CPT-95/P.3184 \end{titlepage}

\newpage {\bf 1. Introduction}\\[6pt]

This paper is written for those who are not willing to become experts
in the field of noncommutative geometry but nevertheless want to
understand the link between this approach and the usual formulation
of the Standard Model of electro-weak interactions. The paper tries
to give simple answers to the following questions: \par
\begin{itemize} \item What has been done in this field so far? \item
What are the several approaches? \item What are the results that are
going to stay and what are those which are, for the moment,
conjectural? \item Is there any relation between this and the
formalism of super-connections (and with super-algebras)?
\end{itemize} We also make a number of comments that may help the
reader to see what is going on in this field.\par The construction of
the full standard model (with usual quarks and leptons but also with
right neutrinos) is carried out by following the simplest possible
route (at least the simplest, from the point of view of the present
authors) and using an appropriate generalization of the notion of
connection. The present paper can be considered as a sequel of
\cite{CEV} but can also be read  independently; it should not be
considered as an expository lecture  on the approach initiated by
\cite{Co1}. \\[6pt]

{\bf 2. The meaning of noncommutative geometry}\\[6pt]

{}From the point of view of Physics, one can summarize the situation
very simply by saying that ``commutative geometry'' is the collection
of mathematical tools describing classical physics whereas ``non
commutative geometry'' is the collection of mathematical tools
describing quantum physics. \par Commutative geometry (or better
``commutative mathematics'') deals with mathematical properties of
spaces (measurable, topological, differentiable, riemannian,
homogeneous...). For the physicist, these "spaces" provide a
mathematical model for the system under study and all the properties
of interest can be expressed in terms of an appropriate class of
(numerical) functions defined on such spaces. It is a fact --
physically obvious but also mathematically rooted -- that properties
of "spaces" are entirely encoded in terms of properties of algebras
of numerical functions (coordinates for example) or of objects
themselves defined from numerical functions (forms, tensors etc.) The
name ``commutative mathematics'' comes from the fact that a set of
numerical functions defined on a space is a commutative (and
associative) algebra for pointwise multiplication and addition of
functions. \par Non commutative geometry (or better ``non commutative
mathematics'') deals with mathematical properties of algebras which
are not necessarily commutative and generalizes -- or tries to
generalize -- the constructions already known for commutative
algebras (i.e. spaces) to non commutative situations (i.e. to
operators). \par This shows that one should maybe not always speak of
``commutative geometry'' or of ``non commutative geometry'' but of
``commutative mathematics'' or of ``non commutative mathematics''.
What we have in mind in the present paper is not the use of non
commutative mathematics in physics (because this could include, among
many other things, the mathematics of quantum statistical mechanics)
but non commutative differential geometry.\par From an
epistemological point of view, and once the concepts of commutative
geometry and/or non commutative geometry have been mathematically
studied, one should probably revert the first general statement of
the present paragraph and define classical physics itself as a human
activity characterized by the wish of understanding what we call
"Nature" in terms of commutative mathematics and define quantum
physics in the same way but where the models are now expressed in
terms of non commutative mathematics. One could even go further and
declare that only the choice of the mathematical model (or models)
gives a meaning (meanings) to the whole thing (Nature) and that there
is no such thing as ``reality'' per se... but we now abandon these
philosophical considerations to return to the differential calculus,
commutative or not.\\[6pt]

{\bf 3. Non commutative versus commutative differential
calculus}\\[6pt]

A branch of non commutative differential geometry is  non commutative
differential calculus. The aim is to be able to consider objects like
$df$ or $\nabla f$, i.e. differentials or covariant differentials,
and to perform computations with them, assuming that $f$ is no longer
a function but an operator acting in some Hilbert space. Such a
calculus has been developed in the recent years. There exist several
kinds of non commutative differential calculi (for instance
\cite{Co2,ACbook,CEV,CHS,DV3,DV4}) and we do not intend here to
describe them all. As a matter of fact, we shall describe none of
them. Indeed, it turns out that a very simple by-product of this
(these) generalization(s) gives us the necessary tools to understand
Higgs fields as generalization of connections (Yang Mills fields). In
some cases, for instance the $U(1)\times U(1)$ model studied  in
\cite{CEV}, this by-product actually belongs to the realm of
commutative geometry because it involves only commutative algebras of
functions on "spaces"! The point is that it was discovered
historically \cite{Co1} only after the new developpements of non
commutative calculus. It is maybe a little bit misleading to call
"non commutative" some of these considerations, first because, in
simple cases, they are not so, and next because they give the reader
the  feeling that he should first master all (or most of) the
niceties of non commutative  differential calculus to understand the
constructions. Of course, this is a matter of taste and some people
 could as well argue that one  should always understand the general
before going to the particular... \\[6pt]

{\bf 4. Commutative non local differential calculus}\\[6pt]

In the previous paragraph, we said that the constructions that are at
the root of our understanding of Higgs field as generalized
connections do not really belong to the realm of non commutative
differential geometry  (they are a by-product). They however
correspond to some commutative -- but non local -- geometry. Let us
see why.\par Consider a discrete set $\{L,R\}$ with two elements that
we call $L$ and $R$. Call $x$ the coordinate function $x(L) \doteq 1,
x(R) \doteq 0$ and  $y$ the coordinate function $y(L) \doteq 0, y(R)
\doteq 1$. Notice that  $xy=yx=0$, $x^2=x, y^2=y$ and $x + y = 1$
where $1$ is the unit function $1(L)=1, 1(R)=1$. An arbitrary element
of the associative (and commutative) algebra ${\cal A}$ generated by
$x$ and $y$ can be written $\lambda x + \mu y$ (where $\lambda$ and
$\mu$ are two complex numbers) and can be represented as a diagonal
matrix $\pmatrix{\lambda & 0 \cr 0 & \mu \cr}$. One can write ${\cal
A} = \CC x \oplus \CC y$ and is isomorphic with $\CC \oplus \CC$. We
now introduce a differential $\delta$ satisfying $\delta^2=0$,
$\delta 1 = 0$ and the usual Leibniz rule, along with formal symbols
$\delta x$ and $\delta y$. It is clear that $\Omega^1$, the space of
differentials of degree $1$ is generated by the two independent
quantities $x\delta x$ and $y\delta y$. Indeed, the relation $x+y=1$
implies $\delta x + \delta y = 0$, the relations $x^2=x$ and $y^2=y$
imply $(\delta x)x+x(\delta x) = (\delta x)$, therefore $(\delta x) x
= (1-x) \delta x$ and $(\delta y) y = (1-y) \delta y$. This implies
also, for example $\delta x = 1 \delta x = x \delta x + y \delta x$,
$x \delta x = - x \delta y$, $y \delta x = (1-x) \delta x$, $(\delta
x)x = y\delta x = - y \delta y$ \etc. More generally, let us call
$\Omega^p$, the space of differentials of degree $p$; the above
relations imply that a base of this vector space is given by
$\{x\delta x\delta x\ldots\delta x,y\delta y\delta y\ldots\delta y
\}$. Call $\Omega^0 = {\cal A} $ and $\Omega = \bigoplus_p \Omega^p
$. This space $\Omega$ is an algebra: We can multiply forms freely
but one of course has to take into account the Leibniz rule, for
instance $x(\delta x) x(\delta x) = x(1-x)(\delta x)^2$. Since each
$\Omega^p$ is two dimensional we can easily represent it in terms of
matrices. More precisely, we can represent the element $\lambda x
(\delta x)^{2p} + \mu y (\delta y)^{2p}$ as the diagonal matrix
$\pmatrix{\lambda & 0 \cr 0 & \mu \cr}$ and the the element $\alpha x
(\delta x)^{2p+1} + \beta y (\delta y)^{2p+1}$ as the off diagonal
matrix $\pmatrix{0 & i\alpha \cr i\beta & 0 \cr}$  In other words we
represent even forms by even (i.e. diagonal) matrices and odd forms
by odd (i.e. off diagonal) matrices; doing so is not only natural but
compulsory if we want the multiplication of matrices to be compatible
with the multiplication in $\Omega$. Indeed, the relations $$
\begin{array}{*2l} x (\delta x)^{2p} x &= x (\delta x)^{2p} \\ x
(\delta x)^{2p} y &= 0 \\ x (\delta x)^{2p+1} x &= 0 \\ x (\delta
x)^{2p+1} y &=x (\delta x)^{2p+1} \end{array} $$ imply that the above
representation using $2 \times 2$ matrices  is indeed a homomorphism
of $\ZZ_2-$graded algebras from the algebra of universal forms
$\Omega$ (graded by the parity of $p$) to the algebra of $2 \times 2$
complex matrices (with $\ZZ_2-$grading associated with the
decomposition of a matrix into diagonal and non diagonal components).
The presence of a factor $i$ in the off diagonal matrices
representing  odd elements (see above expressions) is necessary for
the matrix product to be compatible  with the product in $\Omega$.
Notice that the algebra $\Omega$ is infinite dimensional (since $p$
ranges from $0$ to infinity) but if we represent the whole of
$\Omega$ in terms of $2\times 2$ matrices acting on a fixed
$2$-dimensional vector space, the $p$ grading is lost and only the
$\ZZ_2$ grading is left. The differential $\delta$ obeys the usual
Leibniz rule when it acts on elements of ${\cal A}$ but a graded
Leibniz rule when it acts on elements of $\Omega$, namely
$\delta(\omega_1 \omega_2) = \delta(\omega_1) \omega_2 +
(-1)^{\partial \omega_1} \omega_1 \delta(\omega_2)$ where ${\partial
\omega_1}$ denotes $0$ or $1$ depending if $\omega_1$ is even  or
odd.\par A one-form (this will be interpreted as a Higgs field and
can be seen to define a generalized connection) is an element of
$\Omega^1$. Take $A \doteq ( \varphi \, x \delta x + \bar  \varphi \,
y \delta y)$. The matrix representation of $A$ reads therefore $$ A
=  \pmatrix {0 & i  \varphi \cr i \bar  \varphi & 0} $$ The
corresponding curvature is then $F \doteq \delta A + A^2$, but $A^2 =
- \varphi \bar  \varphi x^2 \, \delta x \delta x - \bar  \varphi
 \varphi \, y^2 \delta y \delta y =  - \varphi \bar  \varphi x \,
\delta x \delta x - \bar  \varphi
 \varphi \, y \delta y \delta y$ and  $\delta A =  \varphi \delta x
\delta x +  \bar  \varphi  \delta y \delta y = (\bar  \varphi +
\varphi)(x \delta x \delta x +  y \delta y \delta y)$, so that the
curvature can also be written
 $$ F =  ( \varphi + \bar  \varphi -  \varphi \bar  \varphi )(x
\delta x \delta x + y \delta y \delta y)
 = \pmatrix { \varphi + \bar  \varphi -  \varphi \bar  \varphi & 0
\cr 0 &  \varphi + \bar  \varphi -  \varphi \bar  \varphi} $$ We now
chose a hermitian product on $\Omega$ by declaring the base $x(\delta
x)^p,y(\delta y)^q$ to be orthonormal. Then $\vert F \vert^2 = \bar F
F = ( \varphi + \bar
 \varphi -  \varphi \bar  \varphi)^2$. One recognizes here a
(shifted) Higgs potential. The previous calculation (expressed in the
language of $K$-cycles) is already discussed in \cite{Co1,Co2} and
can be recognized in \cite{CEV} where it is written in the language
of $2 \times 2$ matrices. \par
 The previous construction could of course be generalized. For
instance, we could take three points rather than two. It is easy to
show that in such a case, $\Omega^1$ is of dimension $6$ and
$\Omega^2$ of dimension $12$. If we take $q$ points the dimension of
$\Omega^p$ is $q(q-1)^p$. More generally, if we take infinitely many
points -- take for instance points belonging to a manifold $X$ --  it
is easy to see that elements of $\Omega^1$ can be defined as
functions $A(x,y)$ of two variables on $X$ such that $A(x,x)=0$ and
that elements of $\Omega^2$ can be defined as functions $F(x,y,z)$ of
three variables on $X$ such that $F(x,y,y) = F(x,x,y) = 0$. \par In
the case of the geometry on the discrete set $\{L,R\}$ -- that is our
main example in the present paper -- we recover the fact that an
element $A$ of $\Omega^1$ considered as a function of two variables
should satisfy the constraints $A(L,L)=A(R,R)= 0$ and can therefore
be written as off-diagonal $2 \times 2$ matrices indexed by $L$ and
$R$. An element $F$ of $\Omega^2$ considered as a function of three
variables should satisfy the constraints
$F(L,L,R)=F(R,R,L)=F(L,R,R)=F(R,L,L)=F(R,R,R)=F(L,L,L)=0$ so that non
zero components are $F(L,R,L)$ and $F(R,L,R)$. The fact that
$dim(\Omega^p) = 2$ for all $p$ explains that we can use a
representation of fixed dimension (namely $2\times 2$ matrices) for
all values of $p$ but one should maybe remember that it would not be
so if we were considering a geometry on more than $2$ points.\par
Notice that we are here doing commutative differential calculus
(because the associative algebra of functions on a set of $2$
elements is just the commutative algebra of diagonal $2$ by $2$
matrices with real or complex entries) but that we are doing a non
local differential calculus because the distance between the two
points labelled $L$ and $R$ can not be made infinitesimally small.
The reader will have recognized that one can interpret the above
results in terms of Higgs fields. This is the subject of our next
section.  \\[6pt]

{\bf 5. What are the Higgses ? The Yukawa interaction term}\\[6pt]

Higgs fields ($ \varphi$) allow left and right fermions ($\psi$) to
communicate. In four dimensional Minkowski space, this is clear from
the trilinear  Yukawa couplings such as $\bar \psi_L  \varphi \psi_R
+ h.c.$ that appear in the Lagrangian density of the Standard Model.
This should be contrasted with terms like $\bar \psi_L \gamma^\mu
A_\mu \psi_L + h.c.$ or $\bar \psi_R \gamma^\mu B_\mu \psi_R + h.c.$
where $A_\mu$ or $B_\mu$ denote usual Yang Mills gauge fields. If we
had no Higgs fields, of course we would have no mass term but also no
possible communication (interaction) between right and left. There
would be no justification for choosing a single connected manifold to
modelize our universe. We would have a Minkowski space-time for the
right movers and a Minkowski space-time for the left movers.
Existence of chirality in four dimensions leads therefore to the
conclusion that we live in two parallel universes, one labelled by
$L$ and the other by $R$. Usual connections -- Yang Mills fields --
connect (infinitesimally) $L$ and $L$ together and $R$ and $R$
together whereas Higgs fields are non local connections that connect
$L$ and $R$ and allow us to identify the two copies of our
universe.\par As explained in all books of particle physics, the
scalar interaction (Yukawa) of quarks is a priori of the form
 $$ {\cal L} =
 (\bar U_L \bar D_L)\pmatrix{ \varphi_+ \cr  \varphi_0 \cr} {\cal
M}_D D_R  +
 (\bar U_L \bar D_L)\pmatrix{\bar  \varphi_0 \cr - \varphi_- \cr}
{\cal M}_U U_R + h.c. $$ where all quark fields of charge $2/3$ are
collected into the multi-spinor field $U = (u \, c \, t)^t$ and
similarly for quarks of charge $-1/3$ with $D = (d \, s \, b)^t.$ The
$3\times 3$ complex matrices ${\cal M}_D$ and ${\cal M}_U$ encode all
the {\cal dimensionless} Yukawa coupling constants (here spinor and
scalar fields have their usual dimensions, namely $3/2$ and $1$ in
units of mass). If we expand the previous expression, we find $$
\begin{array}{*2l} {\cal L} = & (\bar U_L {\cal M}_U \bar  \varphi_0
U_R + \bar U_R {\cal M}_U^\dagger  \varphi_0 U_L) +  (\bar D_L {\cal
M}_D  \varphi_0 D_R + \bar D_R {\cal M}_D^\dagger \bar  \varphi_0
D_L) + \\ & (\bar U_L {\cal M}_D  \varphi_+ D_R + \bar D_R {\cal
M}_D^\dagger  \varphi_- U_L) - (\bar U_R {\cal M}_U^\dagger
\varphi_+ D_L + \bar D_L {\cal M}_U  \varphi_- U_R)  \end{array} $$
Let us now collect {\sl all} left-handed quark fields (of charge
$2/3$ and $-1/3$) of the standard model into a single spinor $\Psi_L
= (U_L D_L)$ and {\sl all} right-handed quark fields into a single
spinor $\Psi_R = (U_R D_R)$. Here $U = (u \, c \, t)$ and $D = (d \,
s \, b)$. The above Yukawa interaction term reads  $$ {\cal L} =
\pmatrix{\bar \Psi_L & \bar \Psi_R \cr} \pmatrix{0 & \Phi \cr
\Phi^\dagger & 0 \cr} \pmatrix{\Psi_L \cr \Psi_R \cr} $$ where  $$
\Phi =  \pmatrix{{\cal M}_U \bar  \varphi_0 & {\cal M}_D  \varphi_+
\cr -{\cal M}_U  \varphi_- & {\cal M}_D  \varphi_0 \cr} $$ The mass
term is obtained by shifting $\varphi_0$ and $\bar  \varphi_0$  by a
real constant with dimension of a mass that we call $\nu / \sqrt 2$.
 The mass term itself is therefore described by the mass matrix $$
{\cal M} = \nu / \sqrt 2  \pmatrix{ {\cal M}_U & 0 \cr 0 & {\cal M}_D
\cr} $$ Writing  $\Psi = (\Psi_L, \Psi_R)^t$, the whole fermionic
lagrangian, for quarks, reads $\bar \Psi \, ({\cal D + A}) \, \Psi$
with  $$
 {\cal D + A} =  \pmatrix{i\gamma^\mu \partial_\mu & {-\cal M} \cr
{-\cal M}^\dagger & i\gamma^\mu \partial_\mu \cr} + \pmatrix
{\gamma^\mu L_\mu &  \Phi \cr  \Phi^\dagger & \gamma^\mu R_\mu \cr}
 $$ Where $L_\mu$ and $R_\mu$ collectively refer to those components
of the gauge fields coupled to the left and right handed sectors and
$\Phi$ collectively refers as before to Higgs fields couplings. The
scalar interaction (Yukawa) of leptons is exactly of the same type.
The only possible difference is that, in the minimal Standard Model,
one does not usually add right neutrinos. We shall actually add such
right neutrinos: They will not be coupled to the gauge fields, of
course, but they will give a mass to the different kinds of Dirac
neutrinos and will be also coupled between themselves -- via mass
matrices -- and to the Higgses (and also therefore, in the unitary
gauge, to the longitudinal part of the gauge bosons). Notice that we
do not consider Majorana neutrinos. Introducing right neutrinos, not
only allows us to use the same formalism for quarks and leptons (the
only difference in the Yukawa interaction term is the replacement of
matrices ${\cal M}_D$ and ${\cal M}_U$ by ${\cal M}_E$ and ${\cal
M}_\nu$ respectively) but also, as we shall see later, simplifies our
analysis. The Yukawa interaction for leptons is
 $$ {\cal L} =
 (\bar \nu_L \bar E_L)\pmatrix{ \varphi_+ \cr  \varphi_0 \cr} {\cal
M}_E E_R  +
 (\bar \nu_L \bar E_L)\pmatrix{\bar  \varphi_0 \cr - \varphi_- \cr}
{\cal M}_\nu \nu_R + h.c. $$ where all leptons fields of charge $-1$
are collected into the multi-spinor field $E = (e \mu \tau)^t$ and
similarly for the neutrinos  $\nu = (\nu_e \nu_\mu \nu_\tau)^t.$ The
$3\times 3$ complex matrices ${\cal M}_E$ and ${\cal M}_\nu$ encode
all the Yukawa coupling constants. The whole fermionic lagrangian,
for leptons, reads as before, but with $$ \Phi =  \pmatrix{{\cal
M}_\nu \bar  \varphi_0 & {\cal M}_E  \varphi_+ \cr -{\cal M}_\nu
\varphi_- & {\cal M}_E  \varphi_0 \cr} $$ In the standard model, one
should consider simultaneously not only the three generations of
leptons but also three copies (for color) of the three generations of
quarks. Taking into account -- as above -- the presence of three
kinds of right neutrinos, we get an interaction term $\bar \Psi
({\cal D + A}) \Psi$, with $\Psi = (\Psi_L \Psi_R)$ and where both
$\Psi_L$ and $\Psi_R$ are multi-spinor fields -- they are column
vectors with $24$ components (since $24 = 3 + 3 + 3 (3 + 3)$), each
component being itself a Weyl fermion.\par In the spirit of
noncommutative geometry, one should think of ${\cal D + A}$ as a
generalization of the Dirac operator (it incorporates masses and
Yukawa couplings) coupled to an algebraic connection. It should be
called the Dirac - Yukawa operator. The first piece in this
expression is a generalized differential operator since the mass
matrix ${\cal M}$ appears as the inverse of a quantity encoding a
discrete set of fundamental lengths. The second piece ${\cal A}$ is a
generalized connection: it incorporates both Yang-Mills and Higgs
fields. \\[6pt]

{\bf 6. The bosonic lagrangian}\\[6pt]

The theory of -- usual -- connections explains why $F = dA + A^2$ is
the natural object (curvature) associated with a Yang-Mills field.
The root of the explanation being that the square of the corresponding
covariant differential is a linear object whose expression is
precisely given by the above formula. In the same way, and as
explained (section 4) in a very simple case, the theory of
generalized connections shows that $ \varphi +\bar  \varphi +
\varphi \bar  \varphi$ is the natural object (curvature) associated
with the Higgs field $ \varphi$ introduced in the section 4 and
defined as a non local connection on the discrete set $\{L,R\}$. \par
Now, we do not have a discrete set but a space $X = M_L \cup M_R$
that is the union of space-time for left-handed movers and
space-time for right-handed movers, in other words, we have the
product of Minkowski space by a discrete set of two elements called
$L$ and $R$. The generalized curvature ${\cal F}$ associated with the
generalized connection ${\cal A}$ introduced in the previous
paragraph is  $$ {\cal F} = \pmatrix{ {\cal F}^{LL} & {\cal F}^{LR}
\cr  {\cal F}^{RL} & {\cal F}^{RR} \cr } $$

With  $$ \begin{array}{*2l} {\cal F}^{LL} &= F^{LL} -  ((\Phi +
\Phi^\dagger) \nu /\sqrt 2 + \Phi \Phi^\dagger) \\ {\cal F}^{RR} &=
F^{RR} -  ((\Phi + \Phi^\dagger)\nu /\sqrt 2 + \Phi^\dagger \Phi) \\
{\cal F}^{LR} &= \nabla \Phi + i (L - R) \nu /\sqrt 2 \\ {\cal
F}^{RL} &= \nabla \Phi^\dagger - i (L - R) \nu /\sqrt 2 \\ \end{array}
$$ and $$ \begin{array}{*2l} \nabla \Phi &= \gamma^\mu ( \partial_\mu
\Phi + i( L_\mu \Phi - \Phi R_\mu))\\ L &= \gamma^\mu L_\mu \\ F^{LL}
&= {1 \over 2} \gamma^\mu \gamma^\nu F_{\mu \nu}^L \\ R &= \gamma^\mu
R_\mu \\ F^{RR} &= {1 \over 2} \gamma^\mu \gamma^\nu F_{\mu \nu}^R \\
\end{array} $$ The symbols $F^L$ and $F^R$ denote the usual
curvatures of Yang-Mills fields associated with hermitian fields $L$
and $R$. The expression of matrix elements of ${\cal F}$ given before
is a non trivial consequence of the formalism of non commutative
geometry (or of a non local commutative differential calculus!) and
can here be taken as a  definition. These expressions can indeed be
computed from the theory of general connections (commutative or not).
 The components of the curvature were obtained first  by \cite{Co1}.
Up to different normalization factors and the  presence of spurious
fields, their expression agrees with the one  given just above. This
analysis was later improved in \cite{Co2} (replacement of the
so-called algebra $\Omega$ by $\Omega_D$). A detailed exposition of
the formalism of \cite{Co2} using K-cycles and Dixmier trace can now
be found in several places \cite{ACbook,DKcar,KS}. The matrix
elements of ${\cal F}$ given above were obtained by \cite{CEV,C1} in
a simple way (and using the above notations). Our method is briefly
recalled in one of the ``comments'' of  section 8.\par

Notice that the above expressions for ${\cal F}$ have a dimension of
a mass squared and that, as a consequence, an arbitrary mass scale
$\nu /\sqrt 2$ appears in the formula. Explicitely, the term
 $(\Phi + \Phi^\dagger)\nu /\sqrt 2 + \Phi \Phi^\dagger$ and its
adjoint
 can be computed from the expressions of $\Phi$ given previously,
 both in the quark and leptonic sectors.

Up to a normalization factors (we shall come back later to this
physically important problem) one recognizes that the trace of $\bar
{\cal F} {\cal F}$ is nothing else than the lagrangian describing the
bosonic sector of the standard model: One obtains directly the
expression that usually comes after a shift by $\nu /\sqrt 2$ in the
Higgs fields $ \varphi_0$ and $\bar  \varphi_0$ (see \cite{CEV} for a
discussion of this point).

In a sense, the discussion could stop at this point. Indeed, we have
seen in section 5 how to re-write the Dirac-Yukawa interaction term
of fermions and in this section how to recover the whole bosonic
sector of the Standard Model by treating Yang Mills fields together
with Higgs fields as different components of a generalized
connection. However, there are several claims made in the literature
about possible constraints on the parameters of the lagrangian that
one could obtain thanks to a formalism of non commutative geometry.
Because we want to clarify this point (at least in the present
formalism) we shall continue the discussion a little further.

 The whole discussion comes actually from our understanding of the
notation $Tr{\cal \bar F F}$ that should denote a real number. From
the one hand,  if we decide to introduce, by hand, as many arbitrary
constants in the expansion of this quantity (that gives rise to the
full bosonic lagrangian of the standard model) as gauge invariance
allows, we recover exactly the standard model with the same
(unpredictive) relations as usual, namely $M_H = \nu \sqrt \lambda$,
$M_W = \nu g/2 = g M_H /(2 \sqrt \lambda)$ and $M_Z = M_W / \cos
\theta$ where $g$, $\theta$, $\nu$ and $\lambda$ are undetermined.
If, on the other hand, we decide to introduce a ${\sl unique}$
constant $1/g^2$ in front of
 $Tr{\cal \bar F F}$ in order  to normalize simultaneously all the
gauge fields and Higgs fields, we obtain non trivial relations. The
interest of the formalism of non commutative differential geometry is
not, for us, tied up with the existence of such relations; it may be,
however,
 that such relations turn out to acquire, some day, a better status.
For this reason, and also because the reader may be interested, we
shall devote the end of this section to discuss them.

After global multiplication by $1/g^2$, we can rescale gauge fields
as usual by $A \rightarrow g A$ and also the Higgs fields by  $\Phi
\rightarrow g \Phi$.
 Under identification with the usual lagrangian one obtains
immediately $g^2 = \lambda$; this relation is quite natural from a
point of view that identifies gauge fields and Yang Mills fields as
different components of a generalized connection. In that case, the
first general relation giving $M_H$ is not modified but the second
relation becomes
  $M_W = M_H/2$. Moreover, as we shall see below, the value of
$\theta$ also gets constrained.

Rather than writing again in full the well known bosonic lagrangian
of the Standard Model, we shall examine several of the terms, as they
appear here. First of all, notice that one can identify the two sides
of $$ \lambda \{  (\varphi + \bar \varphi){\nu \over \sqrt 2} +
\varphi \bar \varphi \}^2 \equiv \mu^2(\varphi + {\nu \over \sqrt
2})^2 + \lambda (\varphi + {{\nu} \over {\sqrt 2}})^4 - {{\mu^4}
\over {4 \lambda}} $$ provided $\nu^2 = -\mu^2 /\lambda$. The mass
value for the Higgs particle coming  from this usual expression is
$M_H = \nu \sqrt \lambda$. Notice also that the left hand side
contains no additive constant (absence of cosmological term).

In our case, the Higgs potential itself coming from ${1 \over g^2}
Tr({\cal \bar F F})$ reads,  $$V(\Phi) = {2 \over g^2} Tr((\Phi +
\Phi^\dagger)\nu /\sqrt 2 + \Phi \Phi^\dagger)((\Phi + \Phi^\dagger)
\nu /\sqrt 2 + \Phi \Phi^\dagger)^\dagger$$

If we now express $\Phi$ in terms of the component Higgs fields and
in terms of the matrices of Yukawa coupings then remove the factor
$1/g^2$, in front, by rescaling the fields, we see that $V(\Phi)$
contains a term equal to $\nu^2  \varphi_0 \bar  \varphi_0 Tr(M_U
M_U^\dagger + M_D M_D^\dagger)$ but the term $Tr \nabla \Phi \nabla
\Phi^\dagger$ leads to a kinetic term
 for $\phi_0$ equal to $2 \nabla  \varphi_0 \nabla  \varphi_0 (Tr M_U
 M_U^\dagger + Tr M_D M_D^\dagger)$ so that the mass of the Higgs
field does not depend on the mass of fermions and stays undetermined
(remember that $\nu$ is a free parameter). Other authors \cite{KS},
using a different  formalism find quite stringent constraints
relating $M_H$ to the fermionic masses. \par The full bosonic
interaction contains also a term ${{\cal F}^{LR}}^\dagger {\cal
F}^{RL}$; using the previous expression for ${{\cal F}^{LR}}$ implies
that the field $L-R$ becomes massive, as it should. Indeed it
corresponds to the $Z$ and $W$ bosons. One may adopt the point of
view that the present formalism  dictates a particular value for the
Weinberg angle; this value turns  out to depend upon the fermionic
content of the theory. Indeed, the gauge fields $L$ and $R$ consist
of three copies of $$ \matrix{ L & = & \frac{i}{\sqrt 2}
\overrightarrow\tau \overrightarrow W & + \frac{i}{\sqrt x}
\normalbaselineskip=18pt \pmatrix{  y  &       \cr
      & y \cr } B  \cr                  R & = & 0  & +
\frac{i}{\sqrt x}  \normalbaselineskip=18pt \pmatrix{  y + 1
&         \cr
      & y - 1 \cr } B } $$ Here $y = 1/3$ for quarks since their weak
hypercharge is equal to
 $ (y=\frac{1}{3},y=\frac{1}{3} \; ;
    y+1= \frac{4}{3},y-1=  - \frac{2}{3})   $ and $y = 0$ for leptons
since their weak hypercharge is equal to  $ (y= - 1, y =   - 1; y+1
=  0 , y-1=  - 2) .$ We are introducing here right neutrinos that are
isospin singlets and for which $y = 0.$

For colourless quarks alone, the
 normalization $\biggl(  \frac{1}{\sqrt x} \biggr)^2 \Biggl[
 \frac{1}{3}^2 +  \frac{1}{3}^2 + \frac{4}{3}^2 +  \left( -
\frac{2}{3} \right)^2 \Biggr] = 1$

\medskip

\noindent would lead to  $x = 22 / 9$ and $tg \theta = 3 / \sqrt 11 $

\bigskip

For leptons alone, the normalization $\biggl( \frac{1}{\sqrt x}
\biggr)^2 \Bigl[ (- 1)^2 + (- 1)^2 + 0^2 + (- 2)^2 \Bigr] = 1$

\noindent would lead to $x = 6$ and $tg \theta = 1 / \sqrt 3 $

\medskip

More generally, if one uses an arbitrary representation and normalize
fields $L$ and $R$ to $1$ as above, one finds $$tg^2 \theta = 4 Tr
I_3^2 / Tr Y^2$$ which, in the case of three families of quarks (with
color) and leptons, gives $tg^2 \theta = 3/5$ (or $sin^2\theta =
3/8)$ as it is in the unified $SU(5)$ theory. This would be therefore
the ``predicted'' value for the Weinberg angle.  However, in the
usual approach, and even without $SU(5)$ unification, one would
obtain exactly the same value by postulating that the gauge group is
not an arbitrary group isomorphic with $SU(3)  \times SU(2) \times
U(1)$ but a group {\it {metrically}} isomorphic  with the
${SU(3)\times SU(2) \times U(1) / (\ZZ_2 \times \ZZ_3)}$  subgroup of
$SU(5)$.  In absence of a principle based on the ideas of group
symmmetries (or a generalization of such a principle), one could
then ask on which grounds one should postulate such a property. The
same argument (or objection) holds here. Indeed gauge invariance
alone allows for the introduction of arbitrary constants in front of
the individual components of the gauge group. The conclusion is
therefore that, although the value $tg^2\theta = 3/5$ appears quite
``naturally'' in this formalism, it should not be taken as an
unescapable consequence of the construction.\par A last possible
``constraint'' concerns the mass of the $W$ (or  $Z$) particle.
Indeed, from the expression of ${\cal F}$ we obtain a  term ${1\over
4} ({\nu \over \sqrt 2})^2 g^2 Tr(L-R)^2 $ that gives a mass to the
$W$ and the $Z$. The trace itself reads $$ \begin{array}{l} L - R
 =  ( \frac{i}{\sqrt 2} (\tau_1 W_1 + \tau_2 W_2)  +  \frac{i}{\sqrt
2} W_3  \normalbaselineskip=16pt \pmatrix{  1 &   0 \cr 0 & - 1 \cr }
+  \frac{i}{\sqrt x} B  \normalbaselineskip=16pt \pmatrix{  y &     0
\cr
    0 & y \cr })
 -  \frac{i}{\sqrt x} B  \normalbaselineskip=16pt \pmatrix{  y+1
&       0 \cr
    0 & y-1 \cr }                             \\ =   \frac{i}{\sqrt
2} (\tau_1 W_1 + \tau_2 W_2)  +  \frac{i}{\sqrt 2}  \left( W_3 -
\sqrt{ \frac{2}{x}} \, B \right) \normalbaselineskip=16pt \pmatrix{
1 &   0 \cr 0 & - 1 \cr } \end{array} $$ so $$ Tr (L - R)^2
 =  W_1^2 + W_2^2 +  \left( W_3 - \sqrt{ \frac{2}{x}} \, B \right)^2
 =  2 W_+ W_- +  \frac{1}{\cos^2 \theta} \cdot Z^2 $$

This gives the relation $M_W = \nu g/2$ which is well known in the
standard model. In general, we have $M_W = g M_H /( 2 \sqrt \lambda)
$ and this becomes only a constraint  (namely $M_H = 2 M_W$) if we
set $\lambda$ to the ``natural''
 value $\lambda = g^2$ as discussed before.
 One could hope that such
 relations could hold at a scale where the previous value for
$\theta$ is experimentally satisfied (maybe at some grand unification
scale). Notice that  other authors \cite{KS}, using a different
formalism (relying upon the choice of another differential algebra),
obtain another type of relations. Of course, we cannot (and will
not)  pretend that other approaches should, or not, lead to the same
``numerical'' relations. Existence of  constraints such as the above
ones can anyway be  criticized since gauge invariance alone allows us
to multiply
 terms of the bosonic lagrangian by  arbitrary  constants; this
possibility can be related to the choice of particular scalar
products in the space of forms \cite{CES}) and there are no
compelling reasons to set such constants equal to one (although it
may look quite natural in  this formalism).

The main conclusion of this section is that the structure of the
whole  bosonic lagrangian of the Standard Model can be obtained from
the  formalism of non commutative geometry. Whether or not one should
look  for constraints and take them seriously is another matter. Our
opinion is that, before  reaching any conclusion on this line, one
should wait till we have a  full understanding of the fully quantized
field theory in terms of  non commutative geometry. \\[6pt]

{\bf 7. Higgs fields and super-algebras}\\[6pt] The space where
$\Psi$ lives is naturally $\ZZ_2$ graded by $L$ and $R$, \ie $\Psi$
can be decomposed into a left and a right part. Therefore
transformations that map $\Psi$ fields to themselves fall naturally
into $2$ kinds: those mapping $L$ to $L$ (and $R$ to $R$) -- we call
them ``even''-- and those mapping $L$ to $R$ (and conversely) -- we
call them ``odd''. Mathematically speaking, the space of these
transformations can be considered as an associative $\ZZ_2$ graded
matrix  algebra whose corresponding Lie super-algebra is usually
denoted by $GL(p|q)$ where $p$ (resp. $q$) is the number of left Weyl
(resp. right) fermions entering the Lagrangian. The usual Yang-Mills
fields can be decomposed onto the even part whereas the Higgs fields
can be decomposed onto the odd part of this algebra. This is a rather
trivial remark since any Yang-Mills theory (and not only the Standard
Model) defined on an even dimensional space-time can be analysed
along the same lines. Another way to express the same idea is to say
that any Yang Mills theory with $p$ left Weyl fermions and $q$ right
Weyl fermions can be formulated in terms of representation theory of
{\sl some} super Lie algebra posessing a representation on a graded
vector space of dimension $p+q$. In the case of the  Standard Model
(with right neutrinos), and because all the fermionic species are
coupled to the {\sl same}  gauge and Higgs bosons, the matrix
describing this interaction can be  decomposed on a subset of the
generators of $GL(24 \vert 24)$. Since  we have only $4$ gauge bosons
and $4$ Higgs bosons, we need only to  use $8$ generators ($4$ even
and $4$ odd ones); in other words we  only need to use (or to
recognize) the Lie superalgebra  $Sl(2\vert 1)$. The physical
representations of interest (namely leptons, quarks and possibly
right neutrinos) correspond to direct sums of $Sl(2|1)$
representations of dimension $3 = 2+1$, $4 = 2+2$ or $1$. This fact
was actually observed long ago \cite{YN,JTMN} and sometimes perceived
as a kind of ``miracle''; for us, we consider this property as almost
tautological. The emergence of Lie superalgebras could lead people to
think that one should try to enlarge the formalism of gauge theory to
accomodate Lie superalgebras... Such attempts have been investigated
in the past and shown to lead to serious problems and have, in any
case, nothing to do with the Standard Model itself and even less with
the  non commutative geometry presentation of the Standard Model. In
order to stress this point, let us consider the following analogy:
one can observe that Dirac spinors form a representation of the
Clifford algebra (the Dirac algebra of $\gamma$-matrices); this is
well known; as a consequence it is also true that the spinors with
four complex components also provide a representation for the (non
simple) Lie algebra generated by taking commutators of {\sl
arbitrary} products of $\gamma$ matrices; this does not mean that the
lagrangian of quantum electrodynamics should be invariant (globally
or locally) under such transformations. The fact that an algebra
(like the full algebra of $\gamma$ matrices) is not directly related
with an invariance of the lagrangian does not make it useless (the
spin group and its Lie algebra can of course be expressed in terms of
the $\gamma$'s but the Clifford algebra itself is much bigger). Not
all algebras related to the mathematics of a physical model need  to
describe ``invariances'' or ``symmetries''; the fact that they do
not does not make them useless!
 The same thing is also true here for the super-algebra along the
representation of which one can decompose the matrices acting on the
vector space spanned by the multi-component spinor fields $\Psi =
(\Psi_L \Psi_R)$. This useful algebra is spanned by $8$ generators.
The first four are matrices that, in the interaction term of the
lagrangian describing interaction between fermions and gauge bosons,
appear as coefficients of the Yang-Mills fields $W_\pm, W_3$ and $B$;
they are denoted, as usual, by $I_{\pm}, I_3$ and $Y$. The last four
are matrices that appear as coefficients of the Higgs fields $
\varphi_+,  \varphi_-,  \varphi_0$ and $\bar  \varphi_0$; they give
rise (after having added the hermitian conjugate) to the Yukawa and
mass interaction term. We call them $-\Omega_+^\prime$, $-\Omega_-$,
$\Omega_-^\prime$ and  $\Omega_+$. More precisely, consider the
following (block) matrices:
 $$ \begin{array}{rlcrl}  \Omega_+        &= \pmatrix{ \matrix{0 & 0
\cr 0 & 0 \cr} & \matrix{g & 0 \cr 0 & 0 \cr} \cr
                   \matrix{0 & 0 \cr 0 & b \cr} & \matrix{0 & 0 \cr 0
                   & 0 \cr} \cr} & & \Omega^\prime_-  &= \pmatrix{
\matrix{0 & 0 \cr 0 & 0 \cr} & \matrix{0 & 0 \cr 0 & -e \cr} \cr
                   \matrix{a & 0 \cr 0 & 0 \cr} & \matrix{0 & 0 \cr 0
& 0 \cr} \cr}  \\ \Omega_-        &= \pmatrix{ \matrix{0 & 0 \cr 0 &
0 \cr} & \matrix{0 & 0 \cr g & 0 \cr} \cr
                   \matrix{0 & 0 \cr -b & 0 \cr} & \matrix{0 & 0
                   \cr 0 & 0 \cr} \cr}  & & \Omega_+^\prime &=
\pmatrix{ \matrix{0 & 0 \cr 0 & 0 \cr} & \matrix{0 & 0 \cr 0 & e \cr}
\cr
                   \matrix{0 & a \cr 0 & 0 \cr} & \matrix{0 & 0 \cr 0
& 0 \cr} \cr}  \\ \end{array} $$  where $a , b, g, e$ are themselves
square matrices, for example of size $3 \times 3$ if we consider only
quarks coming in $3$ families. In this case, we decide to label the
basis as follows:
 $\Psi = (u , c, t)_L \, (d, s, b)_L \, (u, c, t)_R \, (d, s, b,)_R$.
Let us define $\sqrt 2 \, I_\pm = \{\Omega_\pm,\Omega_\pm^\prime\}$,
$2 I_3 = \{\Omega_+,\Omega_-^\prime\} - \{\Omega_-,\Omega_+^\prime\}$
and $Y = \{\Omega_+,\Omega_-^\prime\} +
\{\Omega_-,\Omega_+^\prime\}$.  The electric charge is $Q \doteq I_3
+ Y/2 = {\Omega_+,  \Omega_-^\prime}$ Then, provided matrices
$e,b,g,a$ satisfy the relation $e b + g a = 1$, one can show (it is
straightforward but cumbersome) that the $\Omega$ matrices satisfy
the relations
  $$ \matrix { [I_3,I_\pm] = \pm I_\pm & [I_+,I_-]=I_3   &
[Y,I_\pm]=[Y,I_3]=0  \cr [Y,\Omega_\pm]=-\Omega_\pm &
[Y,\Omega_\pm^\prime] = \Omega_\pm^\prime & [I_3,\Omega_\pm]=\pm {1
\over 2}\Omega_\pm \cr [I_3,\Omega_\pm^\prime] = \pm {1 \over 2}
\Omega_\pm^\prime  & [I_\pm,\Omega_\mp]={1 \over \sqrt 2} \Omega_\pm &
[I_\pm,\Omega_\mp^\prime]={-1 \over \sqrt 2}  \Omega_\pm^\prime \cr
[I_\pm, \Omega_\pm] = [I_\pm, \Omega_\pm^\prime] = 0 &
\{\Omega_\pm,\Omega_\pm \} = \{\Omega_\pm,\Omega_\mp \}   =0 &
\{\Omega_\pm^\prime,\Omega_\pm^\prime \} =
\{\Omega_\pm^\prime,\Omega_\mp^\prime \}  =0 \cr
 \{\Omega_\pm,\Omega_\pm^\prime \} = \sqrt 2 I_\pm &
\{\Omega_\pm,\Omega_\mp^\prime \} = \pm I_3 + Y/2  }$$ One recognizes
here the usual relations defining the Lie super  algebra of
$SL(2|1)$. In the case of quarks, one furthermore impose the
following constraints for the hypercharge generator: $Y =( Y_L^U
Y_L^D Y_R^U Y_R^D )$ with $ Y_L^U =  Y_L^D = 1/3 \id_{3\times 3}$,
$Y_R^U = 4/3 \id_{3\times 3}$ and $Y_R^D = -2/3 \id_{3\times 3}.$
These constraints are satisfied if and only if, on top of the
relation $e b + g a = \id$, the matrices $e,b,g,a$ satisfy also the
relations $-e b + g a = (1/3)\id$, $ 2 a g = (4/3)\id$ and $- 2 b e =
- (2/3)\id.$ Indeed, one finds $Y_L^U = Y_L^D = - e b+g a$, $Y_R^U =
2 a g$ and $Y_R^D = - 2 b e$. This imply in particular $g a = a g$
and $e b = b e$.
 One can then check that matrices $I_\pm$ and $I_3$ are then
automatically what they should be.

One may notice that the above expressions for Omega matrices
describing the gauge and Yukawa couplings of the quark family define
a Lie superalgebra representation which is equivalent to the sum of
(three) irreducible representations (each irreducible itself splits
into the direct sum of a doublet and two singlets under the branching
to the Lie algebra of $SU(2) \times U(1)$).

\par Define now
 $\phi = \bar  \varphi_0 \Omega_+ +  \varphi_0 \Omega_-^\prime -
\varphi_+ \Omega_+^\prime -  \varphi_- \Omega_-$ and write ${\cal C}
= \bar \Psi \phi \Psi.$ This expression can not be real, indeed
${\cal C} = {\cal C}^\dagger$ would imply $g=a^\dagger$,
$b=-e^\dagger$, but the other constraints would lead to a
contradiction ($e e^\dagger = -(1/3)\id $). To obtain a real
expression, one has to add ${\cal C}$ and $ {\cal C}^\dagger$.
Writing  ${\cal L} = {\cal C} + {\cal C}^\dagger$ gives

$$ \begin{array}{*2l} {\cal L} = & (\bar U_L (g + a^\dagger) \bar
\varphi_0 U_R + \bar U_R (g^\dagger + a)  \varphi_0 U_L) +  (\bar D_L
(- e + b^\dagger)  \varphi_0 D_R + \bar D_R (- e^\dagger + b) \bar
\varphi_0 D_L) - \\ & (\bar U_L (e - b^\dagger)   \varphi_+ D_R -
\bar D_R (e^\dagger - b)  \varphi_- U_L) - (\bar U_R (g^\dagger + a)
\varphi_+ D_L - \bar D_L (g + a^\dagger)  \varphi_- U_R)  \end{array}
$$

and we recognize the expression of ${\cal L}$ given in section 5,
with the identification  ${\cal M}_U = g + a^\dagger$ and ${\cal M}_D
= -e + b^\dagger$. Warning: The matrix $\phi$ defined previously in
terms of the $\Omega$ matrices is not equal to the matrix $\Phi$
defined in section 5; in order to compare the two expressions, one
has to first add the conjugated expressions ${\cal C}$ and $ {\cal
C}^\dagger$. Taking into account the constraints on blocks $a,g,e$
and $b$, one obtains the relations: ${\cal M}_U = g + (2/3)
(g^{-1})^\dagger$ and ${\cal M}_D = -e + (1/3) (e^{-1})^\dagger.$
These relations do not imply any ``new'' constraints on mass matrices
${\cal M}_U$ and ${\cal M}_D$ since $g$ and $e$ are themselves
arbitrary. The main interest of those formulae is to provided a new
parametrisation for mass matrices or matrices of Yukawa couplings.
This could, in turn, suggest new phenomenological ansatz for them and
may even give us more insight into the structure of fermionic mass
matrices. Such an ansatz was analysed in \cite{CES} in the case of
two families and  leads to a phenomenological expression of the
Cabibbo angle in terms  of quark masses; another ansatz for matrices
$a$ and $b$ was analysed later in \cite{HS} for the case of three
families. \par

Remark: The quantity ${\cal C}$ may be thought as the contribution to
the lagrangian of a particular representation of $Sl(2|1)$. One can
think of ${\cal C}^\dagger$ as the contribution of the antiquark
representation to the lagrangian. However this identification is a
little bit tricky and may lead to possible mistakes of
interpretation; indeed,  ${\cal C}$ is not hermitian but ${\cal
C}^\dagger$ is not the charge  conjugate representation (in any case
weak interactions usualy violate  charge conjugation and one should
not build a lagrangian that would be
 $C$-even !). Given $\Omega_\pm$ and $\Omega^\prime_\pm$ as before,
one can define the following ``hatted'' $\Omega$ matrices: $\hat
\Omega^\prime_- = \Omega_+^\dagger,\hat \Omega^\prime_+ =
\Omega_-^\dagger, \hat \Omega_+ = \Omega_-^{\prime\dagger}$ and $\hat
\Omega_- = \Omega_+^{\prime\dagger}.$ It is then straightforward to
check that these hatted $\Omega$ matrices generate (thanks to the
same commutation relations) matrices $\hat Y$, $\hat I_\pm$ and $\hat
I_3$, with, for example  $\hat Y = Diag(a^\dagger g^\dagger -
b^\dagger e^\dagger, a^\dagger g^\dagger - b^\dagger e^\dagger, 2
g^\dagger a^\dagger, -2 e^\dagger b^\dagger.)$ We obtain in this way
a new representation (the relation $b^\dagger e^\dagger + e^\dagger
b^\dagger = \id$ being automatically satisfied since $eb+be = \id$).
With ${\cal C}$ as before, we can rewrite ${\cal C}^\dagger$ as
${\cal C}^\dagger = \bar \Psi \phi^\dagger \Psi$ and $\phi^\dagger$
as $$\begin{array}{*2l} \phi^\dagger &= \varphi_0 \Omega_+^\dagger +
\bar  \varphi_0 \Omega_-^{\prime \dagger} -  \varphi_-
\Omega_+^{\prime \dagger} -  \varphi_+ \Omega_-^\dagger \\ &=
\varphi_0 \hat \Omega_-^\prime + \bar  \varphi_0 \hat \Omega_+ -
\varphi_- \hat \Omega_- -  \varphi_+ \hat \Omega_+^{\prime \dagger}
\end{array} $$  so that $\phi^\dagger$ itself appears as the
contribution associated with the ``hatted'' representation. If one
wishes to use
 ${\cal C}^\dagger$ in terms of a contribution of antiparticles,
 for instance $\bar d_R  s_L$ as $-\bar s_R^c d_L^c$, one can do it,
modulo proper  care, but it may be misleading.

\smallskip

For leptons, the idea is the same as for the quarks and, in order to
straigthen even more the analogy, we add right neutrinos to the
Standard Model (they will turn out to be iso-singlets, as they should
be). We shall order the basis as follows:
 $\Psi = (\Psi_L \Psi_R)$ with $\Psi_L =  (\nu_e , \nu_\mu,
\nu_\tau)_L \, (e, \mu, \tau)_L$,
 $\Psi_R =  (\nu_e , \nu_\mu, \nu_\tau)_R \, (e, \mu, ,\tau)_R$ and
define matrices Omega as previously, in terms of new $3 \times 3$
block matrices $e,b,g,a$. However, in the case of leptons, the
constraints for the hypercharge generator are different. Indeed,  $Y
=( Y_L^\nu Y_L^E Y_R^\nu Y_R^E )$ with $Y_L^\nu = Y_L^E = - 1 $,
$Y_R^\nu = 0$ and $Y_R^E = -2.$ These constraints are satisfied if
and only if, on top of the relation $e b + g a = \id $ (which ensures
that commutation relations for $SL(2|1)$ hold), the matrices
$e,b,g,a$ satisfy also the relations $ eb = \id$ and $ag = 0$. With
these constraints, one can then check that matrices $I_\pm$, $I_3$
and $Y$ defined as before in terms of the matrices $\Omega$ are then
automatically what they should be.

One may notice that the above expressions for Omega matrices
describing the gauge and Yukawa couplings of the lepton family
(including right neutrinos) define a Lie superalgebra representation
which is equivalent to the sum of (three) reducible indecomposable
representations (each of them splits into the direct sum of a
doublet, a singlet, and the trivial representation under the
branching to the Lie algebra of $SU(2) \times U(1)$).

 Again, we define
 $\phi = \bar  \varphi_0 \Omega_+ +  \varphi_0 \Omega_-^\prime -
\varphi_+ \Omega_+^\prime -  \varphi_- \Omega_-$ and write ${\cal C}
= \bar \Psi \phi \Psi.$ This expression can not be real, and, in
order to obtain a real expression, one has to add, as before, ${\cal
C}$ and $ {\cal C}^\dagger$. Writing  ${\cal L} = {\cal C} + {\cal
C}^\dagger$ gives  $$ \begin{array}{*2l} {\cal L} = & (\bar \nu_L (g
+ a^\dagger) \bar  \varphi_0 \nu_R + \bar \nu_R (g^\dagger + a)
\varphi_0 \nu_L) +  (\bar E_L (- e + b^\dagger)  \varphi_0 E_R + \bar
E_R (- e^\dagger + b) \bar  \varphi_0 E_L) - \\ & (\bar \nu_L (e -
b^\dagger)   \varphi_+ E_R - \bar E_R (e^\dagger - b)  \varphi_-
\nu_L) - (\bar \nu_R (g^\dagger + a)  \varphi_+ E_L - \bar E_L (g +
a^\dagger)  \varphi_- \nu_R)  \end{array} $$

We recognize the expression of ${\cal L}$ given in section 5, with the
identification  ${\cal M}_{\nu}= g + a^\dagger$ and ${\cal M}_E = -e
+ b^\dagger$ but  the matrices $a,g,e$ and $b$ are not totally
arbitrary since they should here satisfy the constraints $e b = \id $
and $a g = 0$. These relations do not imply any constraints on mass
matrices ${\cal M}_{\nu}$ and ${\cal M}_E$ but provided a new
parametrisation for them. This parametrization in terms of $a,g,e,g$
may, in turn, suggest new phenomenological ansatz (for instance one
can see what happens if these matrices $a,g,e,g$ have particularly
simple forms). Such ansatz should then be considered as educated
guesses but not as ''predictions''.\par

\smallskip Before ending this section, we would like to notice that
there exists still another interesting family of parametrizations for
matrices $a,g,e$ and $b$. The reader can indeed check that, if we
chose arbitrary ($3 \times 3$) matrices $N_L, N_{UR}, N_{DR}$ and
choose $a,g,e$ and $b$ in such a way that $2 g a = (4/3)\id + N_L$,
$2 a g = (4/3)\id + N_{UR}$, $2 e b = (2/3)\id - N_L$ and $2 b e =
(2/3)\id - N_{DR}$, then, all commutation relations for $\Omega$
matrices are still satisfied. The generators $I_3, I_+$ and $I_-$
obtained from them are also equal to what they should be. However,
the obtained hypercharge generator $Y$ is not diagonal (and not
necessarily hermitian) but equal to $(1/3)\id + N_L, (1/3)\id + N_L,
(4/3)\id + N_{UR}, -(2/3)\id + N_{DR}$. In other words, this
describes a family of quarks-like objects which are not eigenstates
of hypercharge (hence of charge). The Lie superalgebra specialist may
relate this possibility to the existence of reducible indecomposable
representations of $SL(2|1)$ whith non diagonal Cartan subalgebra
\cite{Marcu} (take  $N_L, N_{UR}, N_{DR}$ nilpotent matrices).
Relation between family mixing and existence of such representations
was suggested in \cite{CES} but was leading to difficulties
(emergence of flavour changing neutral currents in the quark sector)
that could only be cured by a rather {\it ad hoc} treatment of the
definition of charge conjugacy. Here, we just notice that, after
having defined $\phi$ and ${\cal C}$ as before and added the (usual)
complex conjugate, one obtain a real expression and one can choose to
diagonalize simultaneously $I_3$ and $Y$. The rotated quark-like
objects become now hypercharge (and charge) eigenstates, but the
values of their charges are not standard and deviate from their usual
values by corrections encoded in matrices $N_L, N_{UR}, N_{DR}$. This
last family of parametrization leads therefore to something that
deviates from the Standard Model and we shall not elaborate more on
this topic.

\smallskip The $\ZZ_2$-graded algebra discussed in this section is
not usually mentionned in textbooks explaining the construction of
the Standard Model. However, if one decides to rewrite the lagrangian
in terms of multicomponent spinor fields $\Psi = (\Psi_L \Psi_R)$
gathering all left and right fermionic species in this way, this
algebra (or better representations of it)  appears naturally. It
plays a role very similar to the (Clifford) Dirac algebra itself. We
suggest to call it the ``Yukawa algebra''. Again, one should not
consider this algebra as a ``symmetry'' of  the model and it is
probably better to avoid the word ``symmetry'' in this  context in
order to avoid possible misunderstandings. \\[3pt]

{\bf 8. Comments}\\[3pt] \begin{itemize} \item The roads towards
noncommutative geometry and the standard model: All approaches use
three ingredients: an associative algebra $A$ describing ``space'', a
module over this algebra describing ``matter'' and, finally, a
differential algebra where generalized differentials live. In the
very first papers on the subject, A. Connes uses the algebra $\Omega
A$ of universal forms. He then introduced \cite{Co2} one of its
quotients, called it $\Omega_D A$ and used it in the sequel; this
approach was followed in particular by \cite{DKcar,KS,SI,VG}. In
\cite{DV3,DV30} a differential algebra $\Omega_{Der} A$ was built
from $A$-valued derivations of the algebra $A$ and subsequently used
in various papers \cite{DV4,DV5,DV7}.
 The formalism of A. Connes is very general and should be able to
handle many problems of differential calculus in non commutative
geometry. The formalism proposed in \cite{CEV} is not that general,
but if the purpose is to consider Yang-Mills fields and Higgs fields
as different components of the same mathematical object (an algebraic
connection), and to recover the lagrangian of the standard model in a
very simple way, this formalism is sufficient and does not require
any mathematics beyond an elementary knowledge of $n\times n$
matrices. The remark at the origin of \cite{CEV,CES,C1} is that it is
not necessary to use a $\ZZ$-graded differential algebra to perform
the analysis.  The situation is similar to what happens with ordinary
differential forms: one studies usually the theory of connections
(Yang-Mills fields) and covariant derivatives in terms of (usual)
differential forms like $A_\mu dx^\mu$; however, in order to write a
lagrangian describing the interactions with spinor fields, it is
enough to represent these forms (in particular Yang Mills potential
and curvature) in terms of $\gamma$ matrices  like $A_\mu \gamma^\mu$
(which build a $\ZZ_2$-graded algebra). The same thing is true here:
rather than using a $\ZZ$-graded algebra of generalized differential
forms, we only use its representation on the $\ZZ_2$ graded vector
spanned by left and right spinor fields. This is why we only use
$\ZZ_2$ graded algebras in \cite{CEV} (not $\ZZ$ graded ones). We
find this approach more familiar to physicists, and rich enough to
illustrate the idea of considering  Higgs fields as gauge bosons
associated with the gauging of discrete  directions (namely the jump
between left and right chiralities). \item The general Connes'
formalism involves three steps: One starts with a piece of data
containing the (generalized) Dirac operator ${\cal D}$ -- actually
the Dirac-Yukawa  operator-- acting on a Hilbert space  ${\cal H}$,
together with an  associative algebra ${\cal A}$ describing ``space''
and also acting  on ${\cal H}$. In the case of a pure abelian theory,
${\cal A}$ would  just be  equal to the algebra of complex functions
on space-time. Actually, the formalism of \cite{Co1} takes space-time
as euclidean and the Lorentz signature is recovered only at the end,
thanks to a Wick rotation.
 In the case of a $U(1) \times U(1)$ theory with symmetry breaking,
${\cal A}$ is equal to the direct sum of two algebras of complex
functions on space-time (one his labelled by ``left'' and the other
by ``right''); this algebra is still commutative and can be written
as the space of $2 \times 2$ diagonal matrices with entries that are
functions on space-time. In a more complicated setting where the
gauge group is not abelian, one has just to replace the previous
${\cal A}$ by its tensor product with an appropriate matrix algebra.
In the second step one has to construct a differential algebra
$\Omega_D$ (whose definition relies on the choice of ${\cal D}$) out
of which one  defines the generalized connections and curvatures. The
third step is the construction of the Yang-Mills (or generalized
Yang-Mills) Lagrangian itself and involves the so-called Dixmier
trace as a substitute for integration. The triple $({\cal A, H}, D)$
is called a $K$-cycle or a spectral  triple.  The formalism presented
in \cite{CEV} starts with the same  Dirac-Yukawa operator but does
not require the construction of the algebra $\Omega_D$ and the use of
the Dixmier trace. Its clear advantage is simplicity but it is
lacking the character of generality expressed in \cite{Co2,ACbook}.
The only quantity to be computed is the expression for the
generalized curvature ${\cal F}$ in terms of the generalized
connection ${\cal A}$, but this can be done once and for all. \item
In order to build any example of non commutative geometry, one needs
an associative algebra (replacing ``space''), an algebra of
generalized forms and a module (the space of matter fields). However,
for a given ``space'' and a given kind of matter fields, the choice
of the differential algebra $\Omega$ is not unique (all possible
choices can ultimately seen as quotients of a so-called ``universal''
one). The choice used by A.Connes in \cite{Co1} is not the same as
the choice made by the same author in \cite{Co2} and none of them
coincide either with the differential algebra introduced by
\cite{DV3} or with ours. The choice described in \cite{CHS} is a
differential algebra $\Xi$ equal to the tensor product of usual
differential forms times the differential algebra built in section 4
for the space with $2$ elements $\{L,R\}$. One takes ${\cal A}$ as
$1$-form and, using an appropriate $d$-operator defines ${\cal F}$ as
the $2$-form $d{\cal A} + {\cal A} {\cal A}$. The expression
previously given at the beginning of section 6 is nothing else that
the representation of ${\cal F}$ in the fermionic space. The
calculations can be done very simply at the representation level
(this is recalled in section 8) and this why we skip the discussion
concerning the actual choice of  the $\ZZ$ graded differential
algebra. Our point of view is that this freedom in the choice of the
differential algebra does not matter (in the case of the physical
system studied here) because all such choices lead {\sl essentially}
to the same result, namely to the expression of ${\cal F}$ similar to
the one given in section $6$ and to the usual lagrangian of the
standard model. {\sl ``Essentially''} means that all the results
agree, up to factors of normalization. For instance, the expression
of the Higgs potential is exactly the same but the overall
coefficient may change. For instance, in the Connes' approach (where
one has to divide $\Omega$ by ``junk forms'' to obtain $\Omega_D$),
the kinetic term $\bar{D \varphi} D \varphi$ is proportional to
tr$(MM^{\dagger})$ whereas  the Higgs potential is  proportional to
tr$[((MM^{\dagger})_{\perp})^2]$, with $ MM^{\dagger} )_{\perp}=
MM^{\dagger}-\frac{1}{n} tr \, MM^{\dagger},$ where $M$ is a $n
\times n$ fermionic mass matrix. This potential vanishes whenever
 $MM^{\dagger}$ is  proportional to the unit matrix. In our case,
this was not so (see section 6). The coefficients appearing in front
of the kinetic term for scalar  fields and of the Yukawa potential
depend upon the specific choice of the differential  algebra. Spaces
$\Omega_D^0 (A)$, $\Omega_D^1 (A)$ and $\Omega_D^2 (A)$ for $A = C(M)
\oplus C(M)$ have been computed in \cite{Co2} and it happens that
(when  $MM^\dagger \neq \id$) they are respectively isomorphic, as
vector spaces, to the $\Xi^0$, $\Xi^1$ and $\Xi^2$ described in
\cite{CHS} but the algebraic structure differs. More generally, the
structure of  $\Omega_D$, when $A$ is an arbitrary tensor product of
algebras was investigated in \cite{KT,KPPW,PPS}. Introducing an extra
arbitrary constant in front of the whole  potential amounts to
disregard possible mass  relations. In such a case the several
approaches become completely equivalent, as far as physics  and the
standard model of electroweak interactions are concerned.

\item Calculation of the curvature: In \cite{CEV,CES} the calculation
of the coefficients of the generalized curvature is performed in the
representation space (where only the $\ZZ_2$-grading matters). The
whole $\ZZ$-graded differential algebra $\Omega A$ of universal forms
is therefore mapped onto an algebra that can be taken as the
$Z_2$-graded tensor product of the usual algebra of differential forms
 times another $Z_2$ graded differential algebra built in terms of
even dimensional square matrices. Here we forget about the
$\ZZ$-grading of forms and remember only their parity, or
equivalently, we represent forms by using $\gamma$ matrices belonging
to the Clifford algebra. The product is $(a \otimes B) \odot (a'
\otimes B') \doteq (- 1)^{\partial B \partial a'} (a a') \otimes (B
\land B')$  where $a$ is a $ 2n \times 2n$ matrix and $B$ is a form.
The differential is $$ d (a \otimes B) \doteq d a \otimes B + (-
1)^{\partial a} a \otimes d B $$ If we take ${\cal A}$ as in section
5 and define ${\cal F} \doteq d{\cal A} + {\cal A} \odot {\cal A}$,
we obtain the curvature ${\cal F}$ given at the beginning of section
6. The reader should consult \cite{CEV} for a simple exposition of
this calculation. It is not necessary to use the formalism of Lie
super algebras to obtain these results but one may also very well
choose to use it.
 The definition and calculation of ${\cal A}$ and ${\cal F}$  in the
Connes' formalism is given in \cite{Co2,ACbook,DKcar}.

\item The reader remembers that, in case of spontaneous symmetry
breaking, nature (or the formalism!) chooses a vacuum and that there
are usually infinitely many equivalenty such vacua. Since the
formalism of non commutative geometry brings us a formalism where the
translation in the Higgs field is done automatically, one may wonder
about what has happened with the previous freedom for choosing a
vacuum. The answer is that the freedom of choice is encoded in the
definition of the differential $\delta$. This $\delta$ is not unique
and it can be seen that different choices for it amount to choose
different minima for the Higgs potential (see \cite{CEV}).

\item In the formalism used by \cite{Co2}, the algebra of quaternions
plays a very special role (it acts in particular on the algebra of
generalized differential forms). It may help the reader to notice
that quaternions are already present in the usual formalism of the
standard model (also in ours!) but this fact is not necessarily used
or recognized. The simplest way to see quaternions acting here is to
look at the Yukawa interaction term given in section 5 or,
equivalently, to put together in a square $2\times 2$ matrix the two
Higgs doublets coupling the left fermionic doublets to the fermionic
right and left singlets. Discarding Yukawa coupling constants we have
$$ \Phi =  \pmatrix{ \bar
 \varphi_0 &   \varphi_+ \cr -  \varphi_- &   \varphi_0 \cr} $$ One
recognizes here a quaternion. Indeed, using Pauli matrices
$\sigma_i$, set $\gamma_1 = i\sigma_1$, $\gamma_2 = i\sigma_2$,
$\gamma_3 = - i\sigma_3$, then $\gamma_i^2 = -1$ and $\gamma_i
\gamma_j = \epsilon_{ijk} \gamma_k$. Moreover $$a_0 \id + a_1
\gamma_1+ a_2 \gamma_2+ a_3 \gamma_3 =  \pmatrix{a_0 - i a_3 & i a_1
+ a_2 \cr i a_1 - a_2 & a_0 + i a_3 \cr}$$ This expression can
obviously be identified with $\Phi$.

\item On the nature and value of the constants appearing in the
standard model: In any renormalizable quantum field theory, some
parameters appear in the classical lagrangian. These constants may be
free or may be related by some kind of gauge invariance property that
one wants to enforce at the renormalized level. The free parameters
have to be then fixed by an (arbitrary) renormalization prescription.
We want now to stress the following. First of all,  since these
parameters are  free,  it is obvious that any kind of new constraint
will be superimposed to the usual formalism. To be made widely
acceptable, such a constraint should satisfy two criteria. The first
is that it  should physically work, the next is that it should come
from a kind of aesthetical construction (usually encompassing or
generalizing an inherited formalism).

Our next comment concerns stability by renormalization. It has been
shown \cite{VG2} that some proposed constraints among masses of
particles of the standard model may not be stable with respect to the
renormalization group.  This comment should be properly understood
and maybe taken with a grain of salt. Indeed, the free parameters of
the standard model are... free. Therefore one can renormalize them at
will (at a given scale) and one can, in particular renormalize them
in such a way that any relation between them is satisfied. Of course
it is absolutely true that a renormalization prescription involves
the choice of a scale (this ``substraction point''  is usually chosen
equal to some value of $q^2$ where $q$ is a four-momentum) and that a
numerical relation between renormalized parameters may be deformed by
a change of scale if the relation is not invariant under the
renormalization group. But this fact does not mean that the relation
itself is physically meaningless. For example, the relation $m_{muon}
= 206 \, m_{electron}$  which is valid in the on shell
renormalization scheme of quantum electrodynamics can be imposed and
is actually imposed (because it is experimenally true on shell).
However this last relation is not invariant under the renormalization
group equations of QED! The third and last comment to be made about
these problems of relations between constants of the standard model
was already made in the text (section 6) but we repeat it here.
Descriptions of the Standard Model based on non commutative  geometry
supplemented by the choice of specific scalar products in  the space
of fields seem to lead, at the classical level, to relations between
the -- otherwise  arbitrary --  constants of the model.  Gauge
invariance alone allows  for more freedom; in absence of a full
description of (spontaneously  broken) quantum gauge field theories
in terms of non commutative  geometry, such constraints should be
considered, in our opinion, as  educated guesses. The reader may
refer to \cite{KS,SI} for a detailed  analysis of these constraints
in the Conne's framework. For us, the true  power of the non
commutative geometry desription of the standard  model (and of
quantum physics in general) is not tied up with the  relevance of
such constraints. \item Right neutrinos, simplicity and non trivial
bundles (projectors). \par In its simplest ``version'', the standard
model does not incorporate right neutrinos. From the point of view of
non commutative geometry and if one restricts oneself to the leptonic
sector (take for instance the example of one family), lacking right
neutrinos is a little bit of a nuisance. Indeed, in such a case, and
in the language of A.Connes \cite{Co1}, the bundle of leptonic
species is non trivial and one needs to introduce a projector in the
formalism, projector whose curvature itself enters the final
expression. In the formalism explained in \cite{CEV}, the same
phenomena appears because our approach (based on the tensorization of
two by two complex matrices by arbitrary ones) leads to even
dimensional square matrices. In order to accomodate an odd number of
Weyl fermions (for instance $e_L, e_R$ and $\nu_L$) one has to embedd
the odd dimensional matrix describing the connection into a even
dimensional one by adding line an columns of zeros. But then action
of the $d$ operator creates non zero entries in such places. The
curvature is then not equal to $F = dA + A^2$ but to $F = p(dA +
A^2)p + p dp dp$ where $p$ projects on the odd dimensional subspace
spanned by the Weyl fermions. The result is, as it should, the usual
standard model without right neutrinos. However, introducing right
neutrinos in the game (like in \cite{CES} and like in section 5 of
the present paper) simplifies considerably the formalism because one
does not have to introduce such a projector. One cannot not say that
``Non commutative geometry predicts that the neutrino has a mass''
but it is clear that, from our perspective, the formalism is much
simpler with a right neutrino than without. Let us remind the reader
that such neutrinos are absolutely compatible with present
experimental data since the $\nu_R$ that one introduces for each
family is not coupled to the (transverse part of the) gauge fields.
Its main interest is to give a mass to the corresponding particle
(hence a Dirac spinor) and to introduce mixing between fermionic
families via a fermionic analogue of the Kobayashi-Maskawa matrix.
Introduction of right neutrinos in the Connes' formalism was recently
investigated in \cite{GB}.

\item In \cite{SI}, the authors raise the question: ``Is any
Connes-Lott model a Yang-Mills-Higgs model and conversely?'' Their
answer is clearly ``No''. However, one should not assimilate the
approach initiated in \cite{Co1} and further discussed and detailed
in the last chapter of \cite{ACbook} with non commutative geometry in
general or even with non commutatively inspired models. Other avenues
are possible. Our belief is that it should be possible to distort
sufficiently enough the formalism proposed initially by \cite{Co1}
(or \cite{CEV}) to accomodate many cases of classical gauge field
theory with symmetry breaking... For instance it was shown in
\cite{CFF2} how to introduce ``symmetries'' into the Higgs fields of
a non commutative model of the kind \cite{Co1} in order to be able to
accomodate scalar fields belonging to representations a priori
forbidden in the initial framework. Such a distortion of the
formalism allows of course the construction of more general models
but at the expense of aesthetics.

\item The formalism introduced in \cite{CEV} could induce one to
think that this approach is a particular case of the formalism of
super connections introduced and discussed in \cite{MQ}. This is
actually not so. The algebraic connections defined by \cite{Co1} are
clearly not super connections since they are degree one forms in a
$\ZZ-graded$ differential algebra. Of course this algebra is also
$\ZZ_2$ graded since forms are even or odd but generalized Yang Mills
potentials are defined here as one-forms and not only as odd forms.
The same is true in our case, see the discussion in \cite{CHS}.
Notice that when the differential algebra is represented on the space
of spinors, the $\ZZ$ grading is lost and only the $\ZZ_2$ grading is
left.  Contrarily to what happens in the standard model, where the
generalized gauge field incorporates only Higgs and usual spin one
gauge fields, an algebraic super connection would also incorporate
other kinds of fields. Of course, one can consider a  connection as
some kind of ``truncated'' super connection,  but such a terminology
becomes then pointless and can bring  confusion.

\item The ubiquitous number $24$ (joke). With three families of
quarks, three colors, three families of leptons including right
neutrinos, we have $24$ elementary right handed Weyl fermions and
$24$ elementary left handed Weyl fermions. The number of arbitrary
parameters of the model is also equal to $24$ (the gauge coupling
constants $g, g^\prime$, the $6$ quarks masses, the $6$ leptonic
masses, the $8 = 2 \times 4$ parameters of the leptonic and quark
mixing matrices, the mass of the $W$ and the mass  of the Higgs). We
shall stop here these numerological remarks and suggest the reader
interested  in the beautiful properties of the number $24$ to dive
into almost any book of arithmetics, modular function theory or
higher dimensional cristallography.

\item In the present formalism, the square of the generalized
curvature gives us the whole bosonic classical lagrangian of the
standard model. One may wander about the possibility of maintaining
this unification at all orders of perturbative Q.F.T. In other words,
is it possible to write the effective lagrangian at one loop (at two
loops \etc) of the theory in terms of the generalized curvature
${\cal F}$ ? Such a possibility is by no means ruled out but has not
been proven yet. If true, such a property would pave the way to the
next (ultimate?) goal of non commutative geometry (as far as physics
of electroweak interactions  are concerned): a non perturbative
formulation of a fully quantized gauge field theory. The completion
of such an ambitious program would really be a truely non commutative
achievement.

\item Final comment. Higgs fields and Yang-Mills fields can now be
thought of as two particular components of the same object ${\cal A}$
and that the square of the corresponding generalized curvature gives
us the bosonic lagrangian of the standard model. This unification is
a little bit like the unification of electric and magnetic fields
taken as independent components of the Faraday tensor $F$. By itself,
such a unification is not a new theory in the sense that it was
``already there''. After all one can very well work with
electromagnetism without using a manifestly covariant formalism!
Also, it does not bring necessarily any new numerical prediction.
However, a unification of that kind is important at the conceptual
level. Moreover it usually leads to generalizations or to new ideas
that could be hardly thought of in a less unified framework.
Unification of Higgs and Yang-Mills fields is, for us the most
important success of non commutative differential geometry in
physics. This success is certainly going to stay. \end{itemize}

{\bf Acknowledgments}\\[3pt] We would like to thank our friends and
colleagues at C.P.T. (in particular G. Esposito Farese) and at the
University of Mainz (in particular R. Haussling and F. Scheck) for
many discussions on those topics. One of us (R.C.) wants to thank the
Erwin Schrodinger Institute, in Vienna, for its support and
hospitality.

 \end{document}